\documentclass[11pt]{article}

\usepackage{amsfonts,amssymb,amsthm,graphicx,rotating,appendix,xcolor,helvet,amsmath,setspace}
\usepackage[ansinew]{inputenc}
\usepackage[english]{babel}

\numberwithin{equation}{section}

\begin{document}

\title{Joint imputation procedures for categorical variables}
\author{H\'el\`ene Chaput \and Guillaume Chauvet \and David Haziza \and Laurianne Salembier \and Julie Solard}
\maketitle

\begin{abstract}
Marginal imputation, which consists of imputing each item requiring imputation separately, is often used in surveys. This type of imputation procedures leads to asymptotically unbiased estimators of simple parameters such as population totals (or means), but tends to distort relationships between variables. As a result, it generally leads to biased estimators of bivariate parameters such as coefficients of correlation or odd-ratios. Household and social surveys typically collect categorical variables, for which missing values are usually handled by nearest-neighbour imputation or random hot-deck imputation. In this paper, we propose a simple random imputation procedure, closely related to random hot-deck imputation, which succeeds in preserving the relationship between categorical variables. Also, a fully efficient version of the latter procedure is proposed. A limited simulation study compares several estimation procedures in terms of relative bias and relative efficiency.
\end{abstract}

\vspace*{.30in}
{\noindent  {\small {\em  Key words: balanced random imputation; coefficient of correlation; categorical variable; fully efficient estimator; joint proportion; odd-ratio; random hot-deck imputation.}}}

\section{Introduction}

Single imputation, which consists of replacing a missing value by an artificial value, is often used in statistical agencies for treating item nonresponse. The main objective of imputation is to reduce the nonresponse bias, which may be appreciable when the respondents and non-respondents differ with respect to the study variables. Achieving an efficient bias reduction relies on the availability of auxiliary information, which is a set of variables observed for all the sample units. Imputation leads to a complete rectangular data file, which is attractive for an analyst since complete data estimation methods may be readily applied to obtain point estimates. In some cases, response flags, indicating the item specific response statuses for each unit, are provided in the imputed data file. In some situations, however, the flags are not provided by statistical agencies. \\

\noindent In household and social surveys, missing values are often handled through donor imputation procedures such as nearest-neighbour imputation or random hot-deck imputation. In this paper, we focus on survey weighted random hot-deck imputation, whereby a missing value is imputed by the value of a respondent (donor) selected at random from the set of respondents with probability proportional to its sampling weight. In practice, survey weighted random hot-deck imputation is generally applied independently within imputation classes, defined on the basis of auxiliary information; the reader is referred to Andridge and Little (2010) for more details on random hot-deck imputation. \\

\noindent Most often, survey statisticians are interested in estimating simple parameters such as population totals, means and marginal proportions. In this case, marginal imputation, which consists of imputing variables separately, leads to asymptotically unbiased estimators, provided that the assumed imputation model is correctly specified (Haziza, 2009). For example, one may use random hot-deck imputation for each variable requiring imputation. However, this type of method tends to distort the relationships between variables. As a result, estimators of parameters measuring the relationship between variables may be severely biased, especially if the nonresponse rates are appreciable. It is thus desirable to develop imputation strategies which succeed in preserving the relationship between categorical variables. For bivariate parameters involving continuous variables, Shao and Wang (2002) proposed a joint random regression imputation procedure and showed that it leads to asymptotically unbiased estimators of correlation coefficients. Chauvet and Haziza (2012) proposed a fully efficient version of the Shao-Wang procedure in the sense that the imputation variance is eliminated or considerably reduced. A different approach for dealing with bivariate parameters was considered in Skinner and Rao (2002), who proposed to first use marginal imputation to fill in the missing values and then to adjust for the bias at the estimation stage. \\

\noindent In household and social surveys however, variables are often categorical so that the methods described above are not directly applicable: rather than dealing with means and correlations, we are interested in marginal and joint proportions. We propose a simple joint random imputation procedure that requires the same amount of information that is needed for random hot-deck imputation, and show that it preserves the relationship between categorical variables in the sense that imputed estimators of the joint proportions are approximately unbiased for their population counterparts.  Also, a balanced version is proposed, for which the imputation variance is virtually eliminated. The balanced procedure leads to efficient and approximately unbiased estimators of joint proportions while being more efficient than random hot-deck imputation if the interest lies in estimating the marginal proportions.

\section{Set-up} \label{Sec:setup}

Consider a finite population $U$ of size $N$. Let $x$ denote a categorical study variable with possible characteristics $k=0,\ldots, K-1$. Similarly, let $y$ denote a categorical study variable with possible characteristics $l=0,\ldots, L-1$. We are interested in estimating $p_{k\bullet} = N^{-1} \sum_{i \in U} 1(x_i=k),$ the marginal proportion of units who possess the characteristic $k$ for $x$; $p_{\bullet l} = N^{-1} \sum_{i \in U} 1(y_i=l),$ the marginal proportion of units who possess the characteristic $l$ for $y$; and $p_{kl} = N^{-1} \sum_{i \in U} 1(x_i=k) 1(y_i=l),$ the joint proportion of units who possess both characteristics $k$ for $x$ and $l$ for $y$. \\

\noindent A sample $s$ of size $n$ is selected from $U$ according to some sampling design $p(\cdot)$. Let $w_i=1/\pi_i$ be the sampling weight attached to unit $i$, where $\pi_i=P(i \in s)$ denotes its first-order inclusion probability in the sample. Complete data estimators of $ p_{k\bullet},$   $p_{\bullet l}$ and  $p_{kl}$ are the Horvitz-Thompson (1952) estimators
    \begin{eqnarray}\label{complete_data}
      \hat{p}_{k\bullet} & = & N^{-1} \sum_{i \in s} w_i 1(x_i=k), \nonumber \\
      \hat{p}_{\bullet l} & = & N^{-1} \sum_{i \in s} w_i 1(y_i=l), \\
      \hat{p}_{kl} & = & N^{-1} \sum_{i \in s} w_i 1(x_i=k) 1(y_i=l)\nonumber .
    \end{eqnarray}
The estimators $\hat{p}_{k\bullet},$  $\hat{p}_{\bullet l}$  and $\hat{p}_{kl}$ are design-unbiased for $ p_{k\bullet},$   $p_{\bullet l}$ and  $p_{kl},$ respectively. That is,
\begin{eqnarray*}
      E_p(\hat{p}_{k\bullet}) & = & p_{k\bullet}, \\
      E_p(\hat{p}_{\bullet l}) & = & p_{\bullet l}, \\
      E_p(\hat{p}_{kl}) & = & p_{kl},
    \end{eqnarray*}
where  $E_p(\cdot)$ denotes the expectation with respect to the sampling design. Alternatively, the denominator $N=\sum_{i \in U}1$  in (\ref{complete_data}) can be estimated by $\widehat {N}=\sum_{i \in s}w_i$, which leads to the so-called Hajek estimators of $ p_{k\bullet},$   $p_{\bullet l}$ and  $p_{kl}$ (Hajek, 1971). For simplicity, we confine to the case of the Horvitz-Thompson estimators given by (\ref{complete_data}). In practice, both $x$ and $y$ are prone to missing values and require some form of imputation. \\

\noindent In this paper, we assume that the data are Missing at Random (MAR) in the sense that the response probabilities are not related to the variables of interest after accounting for auxiliary variables recorded for both respondents and nonrespondents (Rubin, 1976). We assume that the finite population $U$ is partitioned into $G$ imputation classes $U^1,\ldots, U^g, \ldots, U^G$ of size $N^1,\ldots, N^g, \ldots, N^G$, respectively. In class $U^g$, denote by $s^g=s \cap U^g$ the sample members; $s_{rr}^g$ the set of $n_{rr}^g$ respondents to both items $x$ and $y$; $s_{rm}^g$ the set of $n_{rm}^g$ respondents to item $x$ only; $s_{mr}^g$ the set of $n_{mr}^g$ respondents to item $y$ only; $s_{mm}^g$ the set of $n_{mm}^g$ non-respondents to both items. We note $\phi_{i\circ}^g \equiv P(i \in s_{\circ}^g | i \in s)$ for any response/nonresponse pattern $\circ \in \{rr,rm,mr,mm\}$. We assume that a given pattern occurs with the same probability for any unit $i \in s^g$, so that we simplify the notation as $\phi_{i\circ}^g=\phi_{\circ}^g$. Within each class $U^g$, we assume that the units respond independently of one another. In practice, we may ensure that imputation classes satisfy the previous assumptions by first selecting the auxiliary variables that are related to the probability of response to $x$ and $y$ and fitting a polytomous logistic regression model using the selected auxiliary variables as predictors. For sample unit $i$, we obtain the vector of estimated response probabilities $(\hat {\phi}_{irr}, \hat {\phi}_{irm}, \hat {\phi}_{imr}, \hat {\phi}_{imm})'$. Based of these vectors, the sample is then partitioned into classes by using a classification algorithm (e.g., the $k$-means algorithm). This method can be viewed as an extension of the so-called score method (Haziza and Beaumont, 2007) to the case of two study variables. \\

\noindent The population proportions of interest may be rewritten as
    \begin{eqnarray*}
      p_{k\bullet} = N^{-1} \sum_{g=1}^G N^{g}~p_{k\bullet}^g & \textrm{ with } & p_{k\bullet}^g=(N^g)^{-1} \sum_{i \in U^g} 1(x_i=k), \\
      p_{\bullet l} = N^{-1} \sum_{g=1}^G N^{g}~p_{\bullet l}^g & \textrm{ with } & p_{\bullet l}^g=(N^g)^{-1} \sum_{i \in U^g} 1(y_i=l), \\
      p_{kl} = N^{-1} \sum_{g=1}^G N^{g}~p_{kl}^g & \textrm{ with } & p_{kl}^g=(N^g)^{-1} \sum_{i \in U^g} 1(x_i=k) 1(y_i=l).
    \end{eqnarray*}
Similarly, the complete data estimators (\ref{complete_data}) may be rewritten as
    \begin{eqnarray*}
      \hat{p}_{k\bullet} = N^{-1} \sum_{g=1}^G \hat{N}^{g}~\hat{p}_{k\bullet}^g & \textrm{ with } & \hat{p}_{k\bullet}^g=(\hat{N}^g)^{-1} \sum_{i \in s^g} w_i 1(x_i=k), \\
      \hat{p}_{\bullet l} = N^{-1} \sum_{g=1}^G \hat{N}^{g}~\hat{p}_{\bullet l}^g & \textrm{ with } & \hat{p}_{\bullet l}^g=(\hat{N}^g)^{-1} \sum_{i \in s^g} w_i 1(y_i=l), \\
      \hat{p}_{kl} = N^{-1} \sum_{g=1}^G \hat{N}^{g}~\hat{p}_{kl}^g & \textrm{ with } & \hat{p}_{kl}^g=(\hat{N}^g)^{-1} \sum_{i \in s^g} w_i 1(x_i=k) 1(y_i=l),
    \end{eqnarray*}
where $\hat{N}^{g}=\sum_{i \in s^g} w_i$ is an estimator of the $g$-th class size, $N_g$. \\

\noindent Let $x_i^*$ and $y_i^*$ be the imputed values used to replace the missing $x_i$ and $y_i$. Imputed estimators of $p_{k\bullet}$, $p_{\bullet l}$ and  $p_{kl}$  are respectively
    \begin{eqnarray}
      \hat{p}_{k\bullet,I} & = & N^{-1} \sum_{g=1}^G \sum_{i \in s_{r\bullet}^g} w_i 1(x_i=k) + N^{-1} \sum_{g=1}^G \sum_{i \in s_{m \bullet}^g} w_i 1(x_i^*=k), \nonumber \\
      \hat{p}_{\bullet l,I} & = & N^{-1} \sum_{g=1}^G \sum_{i \in s_{\bullet r}^g} w_i 1(y_i=l) + N^{-1} \sum_{g=1}^G \sum_{i \in s_{\bullet m}^g} w_i 1(y_i^*=l), \label{est:imp} \\
      \hat{p}_{kl,I} & = & N^{-1} \sum_{g=1}^G \sum_{i \in s_{rr}^g} w_i 1(x_i=k) 1(y_i=l) + N^{-1} \sum_{g=1}^G \sum_{i \in s_{rm}^g} w_i 1(x_i=k) 1(y_i^*=l) \nonumber \\
                     & + & N^{-1} \sum_{g=1}^G \sum_{i \in s_{mr}^g} w_i 1(x_i^*=k) 1(y_i=l) + N^{-1} \sum_{g=1}^G \sum_{i \in s_{mm}^g} w_i 1(x_i^*=k) 1(y_i^*=l), \nonumber
    \end{eqnarray}
where $s_{r\bullet}^g=s_{rr}^g \cup s_{rm}^g$ denotes the set of respondents to item $x$ in class $g$; $s_{m\bullet}^g=s_{mr}^g \cup s_{mm}^g$ denotes the set of non-respondents to item $x$ in class $g$, and $s_{\bullet r}^g$ and $s_{\bullet m}^g$ corresponding to item $y$ are similarly defined.  Once the data have been imputed, the computation of (\ref{est:imp}) does not require the response flags to be available in the imputed data file. Complete data estimation procedures may thus be readily applied by secondary analysts, which is an important practical aspect. \\

\noindent In order to study the properties of an imputed estimator $\hat{p}_{\diamond,I}$ of a proportion $p_{\diamond}$, we express its total error as
    \begin{eqnarray} \label{eq:3}
    \hat{p}_{\diamond,I}-p_{\diamond} & = & (\hat{p}_{\diamond} -p_{\diamond}) +\left(\tilde{p}_{\diamond,I} -\hat{p}_{\diamond}\right) + \left(\hat{p}_{\diamond,I} -\tilde{p}_{\diamond,I} \right),
    \end{eqnarray}
where $\tilde{p}_{\diamond,I} \equiv E_I\left( \hat{p}_{\diamond,I}\right),$  for $\diamond \in \{k\bullet,\bullet l,kl\}$, and $E_I(\cdot)$ denotes the expectation with respect to the imputation mechanism, conditionally on the sample $s$ and on the sets of respondents to items $x$ and $y$. In other words, $E_I(\cdot)$ denotes the average with respect to the random selection of donors in the case of a random imputation method. The first term on the right hand side of (\ref{eq:3}) represents the sampling error, whereas the second and the third terms represent the non-response error and the imputation error. The imputation error occurs solely from the random imputation mechanism. We seek an imputation procedure under which the non-response bias
    \begin{eqnarray*}
      B_{pqI}(\hat{p}_{\diamond,I}) & \equiv & E_{p} E_{q} E_{I} \left(\hat{p}_{\diamond,I}-\hat{p}_{\diamond}\right) = E_{p} E_q \left(\tilde{p}_{\diamond,I}-\hat{p}_{\diamond}\right)
    \end{eqnarray*}
is approximately equal to $0$, where $E_q(\cdot)$ denotes the expectation with respect to the assumed non-response model, conditionally on the sample $s$. \\

\noindent We focus on survey weighted random hot-deck imputation, which consists of selecting a donor at random from the set of respondents with probability proportional to its sampling weight, and then using the donor's item value(s) to "fill in" for the missing value of a non-respondent. Marginal random hot-deck imputation, which consists of imputing $x$ and $y $ separately, tends to attenuate the relationship between items being imputed.  As a result, this method introduces a bias in the estimation of $p_{kl}$ that may be severe if the non-response rate is appreciable. In practice, it is customary to use a slightly different version of random hot-deck imputation that consists of using a common donor when both $x$ and $y$  are missing. For any class $U^g$, we proceed as follows:
\begin{itemize}
\item [(i)] for $i \in s_{mr}^g$, missing $x_{i}$ is imputed by $x_{i}^{*}=k$ with probability
    \begin{eqnarray} \label{p:kbul:ac}
      \hat{p}_{k\bullet,ac}^g \equiv (\hat{N}_{r\bullet}^g)^{-1} \sum_{i \in s_{r\bullet}^g} w_i 1(x_i=k)
    \end{eqnarray}
estimated from the available cases (AC) in class $g$ for item $x$, and $\hat{N}_{r\bullet}^g= \sum_{i \in s_{r\bullet}^g} w_i$;
\item [(ii)] for $i \in s_{rm}^g$, missing $y_{i}$ is imputed by means of an analogous procedure;
\item [(iii)]for $i \in s_{mm}^g$, missing $\left(x_{i},y_{i}\right)$ is imputed by $\left(x_{i}^{*},y_{i}^{*}\right)=\left(k,l\right)$ with probability
    \begin{eqnarray} \label{p:kl:cc}
    \hat{p}_{kl,cc}^g \equiv (\hat{N}_{rr}^g)^{-1} \sum_{i \in s_{rr}^g} w_i 1(x_i=k) 1(y_i=l)
    \end{eqnarray}
estimated from the complete cases (CC) in class $g$ to items $x$ and $y$, with $\hat{N}_{rr}^g= \sum_{i \in s_{rr}^g} w_i$.
\end{itemize}

\noindent When one variable only is missing, random hot-deck imputation estimates its distribution separately from available cases for this variable. When both variables are missing, their distribution is estimated jointly from complete cases for both variables. Random hot-deck imputation succeeds in preserving the marginal distributions of $x$ and $y$. Therefore, $B_{qI}(\hat{p}_{k\bullet,I}) \simeq  0$ and $B_{qI}(\hat{p}_{\bullet l,I}) \simeq 0$ for any characteristics $k$ and $l.$  Although this imputation procedure generates less bias than marginal random hot-deck imputation, there generally remains some bias when estimating the joint proportions, since
    \begin{eqnarray}
      B_{pqI}(\hat{p}_{kl,I}) \simeq - N^{-1} \sum_{g=1}^G N^g (\phi_{rm}^g+\phi_{mr}^g) (p_{kl}^g-p_{k\bullet}^g p_{\bullet l}^g). \label{BqI:randHD}
    \end{eqnarray}
The proof of (\ref{BqI:randHD}) is given in Appendix \ref{sec:proof:BqI:randHD}. The asymptotic bias vanishes if $\phi_{rm}^g=\phi_{mr}^g=0$ for any $g$, which means that items $x$ are $y$ may not be missing separately, or if both $x$ and $y$ are unrelated within imputation classes.

\section{Proposed imputation procedures} \label{sec:imp:proc}

\noindent To account for the existing relationship between variables, we propose two imputation procedures, where the distribution of $x$ is estimated conditionally on $y$ if $x$ only is missing, and where the distribution of $y$ is estimated conditionally on $x$ if $y$ only is missing. For any unit $i \in U^g$, we note
    \begin{eqnarray*}
      \hat{p}_{k|l,cc}^g & = & \frac{\sum_{i \in s_{rr}^g} w_i 1(x_i=k) 1(y_i=l)}{\sum_{i \in s_{rr}^g} w_i 1(y_i=l)}
    \end{eqnarray*}
the estimated probability that $x_i=k$ when $y_i=l$, and
    \begin{eqnarray*}
      \hat{p}_{l|k,cc}^g & = & \frac{\sum_{i \in s_{rr}^g} w_i 1(x_i=k) 1(y_i=l)}{\sum_{i \in s_{rr}^g} w_i 1(x_i=k)}
    \end{eqnarray*}
the estimated probability that $y_i=l$ when $x_i=k$. \\

\noindent As pointed out by Chauvet et al. (2011) and Chauvet and Haziza (2012), imputing missing values may be performed by sampling within populations of cells, separately for each of the sub-samples $s_{mr}^g$, $s_{rm}^g$ and $s_{mm}^g$. We introduce the following notation: for any integer $q=1,\ldots,K L$, let $k_q$ and $l_q$ be the two integers such that $q=k_q \times L+(l_q+1)$.
\begin{itemize}
\item [(i)] To handle  units in $s_{mr}^g$, we create a population of cells $U_{mr}^{g*}$ of size $n_{mr}^g \times K$. Each cell $(i,k)$ is assigned the probability of selection $\hat{p}_{k|y_i,cc}^g$ and the $K L$-vector of values $\mathbf{t}_{ik}=\left\{(\mathbf{t}_{ik})_1,\ldots,(\mathbf{t}_{ik})_{KL} \right\}^{\top}$ with
        \begin{eqnarray*}
          (\mathbf{t}_{ik})_q & = & w_i~\hat{p}_{k|y_i,cc}^g~1(k=k_q)~1(y_i=l_q).
        \end{eqnarray*}
    A random sample $s_{mr}^{g*}$ of size $n_{mr}^g$ is selected from $U_{mr}^{g*}$, and missing $x_{i}$ is imputed by $x_{i}^{*}=k$ if the cell $(i,k)$ is selected.
\item [(ii)] To handle  units in $s_{rm}^g$, we create a population of cells $U_{rm}^{g*}$ of size $n_{rm}^g \times L$. Each cell $(i,l)$ is assigned the probability of selection $\hat{p}_{l|x_i,cc}^g$ and the $K L$-vector of values $\mathbf{t}_{il}=\left\{(\mathbf{t}_{il})_1,\ldots,(\mathbf{t}_{il})_{KL} \right\}^{\top}$ with
        \begin{eqnarray*}
          (\mathbf{t}_{il})_q & = & w_i~\hat{p}_{l|x_i,cc}^g~1(x_i=k_q)~1(l=l_q).
        \end{eqnarray*}
    A random sample $s_{rm}^{g*}$ of size $n_{rm}^g$ is selected from $U_{rm}^{g*}$, and missing $y_{i}$ is imputed by $y_{i}^{*}=l$ if the cell $(i,l)$ is selected.
\item [(iii)] To handle units in $s_{mm}^g$, we create a population of cells $U_{mm}^{g*}$ of size $n_{mm}^g \times (K L)$. Each cell $(i,q')$ is assigned the probability of selection $\hat{p}_{k_{q'} l_{q'},cc}^g$ and the $K L$-vector of values $\mathbf{t}_{iq'}=\left\{(\mathbf{t}_{iq'})_1,\ldots,(\mathbf{t}_{iq'})_{KL} \right\}^{\top}$ with
        \begin{eqnarray*}
          (\mathbf{t}_{iq'})_q & = & w_i~\hat{p}_{k_{q'} l_{q'},cc}^g~1(k_{q'}=k_q)~1(l_{q'}=l_q).
        \end{eqnarray*}
     A random sample $s_{mm}^{g*}$ of size $n_{mm}^g$ is selected from $U_{mm}^{g*}$, and missing $(x_i,y_{i})$ is imputed by $(x_i^*,y_{i}^{*})=(k_q,l_q)$ if the cell $(i,q)$ is selected.
\end{itemize}

In the populations $U_{mr}^{g*}$, $U_{rm}^{g*}$ and $U_{mm}^{g*}$, each row stands for a non-respondent, and each column for a possible imputed value. We impose that
\begin{itemize}
  \item[C1:] the samples $s_{mr}^{g*}$, $s_{rm}^{g*}$ and $s_{mm}^{g*}$ are drawn so that exactly one cell per row is selected.
\end{itemize}
The constraint C1 is required since exactly one imputed value must be selected for each non-respondent. Imposing only the constraint C1 results in the \textit{joint random hot-deck imputation} procedure which may be alternatively described as follows:
\begin{itemize}
\item [(i)] for $i \in s_{mr}^g$, missing $x_{i}$ is imputed by $x_{i}^{*}=k$ with probability $\hat{p}_{k|y_i,cc}^g$,
\item [(ii)] for $i \in s_{rm}^g$, missing $y_{i}$ is imputed by $y_{i}^{*}=l$ with probability $\hat{p}_{l|x_i,cc}^g$,
\item [(iii)]for $i \in s_{mm}^g$, missing $\left(x_{i},y_{i}\right)$ is imputed by $\left(x_{i}^{*},y_{i}^{*}\right)=\left(k,l\right)$ with probability $\hat{p}_{kl,cc}^g$.
\end{itemize}
It is shown in Appendix \ref{appen:B} that $B_{pqI}(\hat{p}_{\diamond,I}) \simeq 0$ under this imputation procedure, for $\diamond \in \{k\bullet,\bullet l,kl\}$ and any characteristics $k$ and $l$. Guidelines are given in Appendix \ref{sec:extension} to extend the joint random hot-deck imputation procedure to the case of more than two missing items. A drawback of the proposed procedure is that it suffers from an additional variability, called the imputation variance, due to the random selection of donors. To eliminate the imputation variance, we further impose that
\begin{itemize}
  \item[C2:] the samples $s_{mr}^{g*}$, $s_{rm}^{g*}$ and $s_{mm}^{g*}$ are drawn so that the following balancing equations are satisfied:
    \begin{eqnarray}
      \sum_{(i,k) \in s_{mr}^{g*}} \left(\hat{p}_{k|y_i,cc}^g\right)^{-1} \mathbf{t}_{ik} & = & \sum_{(i,k) \in U_{mr}^{g*}} \mathbf{t}_{ik}, \label{bal:eq:1}\\
      \sum_{(i,l) \in s_{rm}^{g*}} \left(\hat{p}_{l|x_i,cc}^g\right)^{-1} \mathbf{t}_{il} & = & \sum_{(i,l) \in U_{rm}^{g*}} \mathbf{t}_{il}, \label{bal:eq:2}\\
      \sum_{(i,q) \in s_{mm}^{g*}} \left(\hat{p}_{k_q l_q,cc}^g\right)^{-1} \mathbf{t}_{iq} & = & \sum_{(i,q) \in U_{mm}^{g*}} \mathbf{t}_{iq}. \label{bal:eq:3}
    \end{eqnarray}
\end{itemize}
If the constraint C2 is exactly satisfied, we prove in Appendix \ref{appen:C} that $\hat{p}_{\diamond,I} - \tilde{p}_{\diamond,I} = 0$ for $\diamond \in \{k\bullet,\bullet l,kl\}$ and any characteristics $k$ and $l$. As a result, the imputation error in (\ref{eq:3}) is equal to zero and the imputation variance vanishes. If both constraints C1 and C2 are imposed in the selection of cells, we obtain the \textit{balanced joint random hot-deck imputation} procedure. The constraints C1 and C2 may be satisfied by selecting the samples $s_{mr}^{g*}$, $s_{rm}^{g*}$ and $s_{mm}^{g*}$ by means of the cube method originally developed in the context of balanced sampling; see Deville and Till\'e (2004) and Chauvet et al. (2011). The extension of the above procedure to the case of three categorical procedures is presented in Appendix C.

\section{Alternative estimators} \label{sec:alter:estim}

In this section, we present some alternative estimation procedures for $ p_{k\bullet},$   $p_{\bullet l}$ and  $p_{kl}.$ In Section 7, these procedures are compared empirically to the methods described in Sections \ref{Sec:setup} and \ref{sec:imp:proc} in terms of bias and relative efficiency. We start by the complete case (CC) estimators
     \begin{eqnarray}
          \hat{p}_{k\bullet,cc} = \hat{N}_{rr}^{-1} \sum_{g=1}^G \hat{N}_{rr}^{g}~\hat{p}_{k\bullet,cc}^g & \textrm{ with } & \hat{p}_{k\bullet,cc}^g=(\hat{N}_{rr}^g)^{-1} \sum_{i \in s_{rr}^g} w_i 1(x_i=k), \nonumber \\
          \hat{p}_{\bullet l,cc} = \hat{N}_{rr}^{-1} \sum_{g=1}^G \hat{N}_{rr}^{g}~\hat{p}_{\bullet l,cc}^g & \textrm{ with } & \hat{p}_{\bullet l,cc}^g=(\hat{N}_{rr}^g)^{-1} \sum_{i \in s_{rr}^g} w_i 1(y_i=l), \label{cc:est}\\
          \hat{p}_{kl,cc} = \hat{N}_{rr}^{-1} \sum_{g=1}^G \hat{N}_{rr}^{g}~\hat{p}_{kl,cc}^g & \textrm{ with } & \hat{p}_{kl,cc}^g=(\hat{N}_{rr}^g)^{-1} \sum_{i \in s_{rr}^g} w_i 1(x_i=k) 1(y_i=l), \nonumber
     \end{eqnarray}
which are based on the responding units to both $x$ and $y$, where $\hat{N}_{rr}=\sum_{g=1}^G \hat{N}_{rr}^g$. The bias of CC estimators can be approximated by
    \begin{eqnarray}
      B_{pqI}(\hat{p}_{k \bullet,cc}) & \simeq & \frac{\sum_{g=1}^G N_g \{\phi_{rr}^g-\bar {\phi}_{rr}\}\{p_{k \bullet}^g-p_{k\bullet}\}}{\sum_{g=1}^G N_g \phi_{rr}^g}, \nonumber \\
      B_{pqI}(\hat{p}_{\bullet l,cc}) & \simeq & \frac{\sum_{g=1}^G N_g \{\phi_{rr}^g-\bar {\phi}_{rr}\}\{p_{\bullet l}^g-p_{\bullet l}\}}{\sum_{g=1}^G N_g \phi_{rr}^g}, \label{bias:cc:est} \\
      B_{pqI}(\hat{p}_{kl,cc}) & \simeq & \frac{\sum_{g=1}^G N_g \{\phi_{rr}^g-\bar {\phi}_{rr}\}\{p_{kl}^g-p_{kl}\}}{\sum_{g=1}^G N_g \phi_{rr}^g}, \nonumber
    \end{eqnarray}
where $\bar {\phi}_{rr}=N^{-1} \sum_{g=1}^G N_g \phi_{rr}^g$. From (\ref{bias:cc:est}), CC estimators are biased if there is an association between the probability of responding to both variables and the proportion we wish to estimate. \\

\noindent The bias of the CC estimators can be removed by accounting for class information. This leads to the adjusted complete case (ACC) estimators
  \begin{eqnarray}
      \hat{p}_{k\bullet,acc} = N^{-1} \sum_{g=1}^G \hat{N}^{g}~\hat{p}_{k\bullet,cc}^g,  \nonumber \\
      \hat{p}_{\bullet l,acc} = N^{-1} \sum_{g=1}^G \hat{N}^{g}~\hat{p}_{\bullet l,cc}^g, \label{acc:est}\\
      \hat{p}_{kl,acc} = N^{-1} \sum_{g=1}^G \hat{N}^{g}~\hat{p}_{kl,cc}^g. \nonumber
    \end{eqnarray}
It can be shown that $B(\hat{p}_{\diamond,acc}) \simeq 0$ for any $\diamond \in \{k\bullet,\bullet l,kl\}$. The ACC estimators may be viewed as propensity score adjusted estimators, where the response probability of a unit in a given imputation class is estimated by the response rate to both items within the same class. However, implementing ACC estimators in order to obtain a complete imputed data file will necessarily lead to "impossible values". For example, in the case of a binary variable (with possible values 0 and 1), the imputed values will never  be equal to either 0 or 1 but will lie in the interval $(0,1),$ which is a drawback from a micro-data point of view. In contrast, the imputation procedures described in Section 3 and 4 use the values of donor to replace the missing values, which eliminates the problem of impossible values. \\

\noindent Another set of estimators are based on available cases, which leads to the available case (AC) estimators
 \begin{eqnarray}\label{ac:est}
 \hat{p}_{k\bullet,ac} & = & \hat{N}_{r\bullet}^{-1} \sum_{g=1}^G \hat{N}_{r\bullet}^g~\hat{p}_{k\bullet,ac}^g \textrm{ with } \hat{p}_{k\bullet,ac}^g=(\hat{N}_{r\bullet}^g)^{-1} \sum_{i \in s_{r\bullet}^g} w_i 1(x_i=k), \nonumber \\
 \hat{p}_{\bullet l,ac} & = & \hat{N}_{\bullet r}^{-1} \sum_{g=1}^G \hat{N}_{\bullet r}^g~\hat{p}_{\bullet l,ac}^g \textrm{ with } \hat{p}_{\bullet l,ac}^g=(\hat{N}_{\bullet r}^g)^{-1} \sum_{i \in s_{\bullet r}^g} w_i 1(y_i=l),  \\
 \hat{p}_{kl,ac} & = & \hat{p}_{kl,cc}, \nonumber
 \end{eqnarray}
where $\hat{N}_{r\bullet}= \sum_{g=1}^G \hat{N}_{r\bullet}^g$, and $\hat{N}_{\bullet r}$ is defined similarly. The bias of AC estimators can be approximated by
    \begin{eqnarray}
      B_{pqI}(\hat{p}_{k \bullet,ac}) & \simeq & \frac{\sum_{g=1}^G N_g \{\phi_{r\bullet}^g-\bar{\phi}_{r\bullet}\}\{p_{k \bullet}^g-p_{k\bullet}\}}{\sum_{g=1}^G N_g \phi_{r\bullet}^g}, \nonumber \\
      B_{pqI}(\hat{p}_{\bullet l,ac}) & \simeq & \frac{\sum_{g=1}^G N_g \{\phi_{\bullet r}^g-\bar{\phi}_{\bullet r}\}\{p_{\bullet l}^g-p_{\bullet l}\}}{\sum_{g=1}^G N_g \phi_{\bullet r}^g}, \label{bias:ac:est} \\
      B_{pqI}(\hat{p}_{kl,ac}) & \simeq & \frac{\sum_{g=1}^G N_g \{\phi_{rr}^g-\bar{\phi}_{rr}\}\{p_{kl}^g-p_{kl}\}}{\sum_{g=1}^G N_g \phi_{rr}^g}, \nonumber
    \end{eqnarray}
where $\phi_{r\bullet}^g=\phi_{rr}^g+\phi_{rm}^g$ and $\bar{\phi}_{r\bullet}=N^{-1} \sum_{g=1}^G N_g \phi_{r\bullet}^g$; $\phi_{\bullet r}^g$ and $\bar{\phi}_{\bullet r}$ are defined similarly. An AC estimator is thus biased if there exists an association between the probability of responding to the required variables and the proportion we wish to estimate. \\

\noindent The bias can be removed by accounting for class information, which leads to the adjusted available case (AAC) estimators
  \begin{eqnarray}
      \hat{p}_{k\bullet,aac} = N^{-1} \sum_{g=1}^G \hat{N}^{g}~\hat{p}_{k\bullet,ac}^g,  \nonumber \\
      \hat{p}_{\bullet l,aac} = N^{-1} \sum_{g=1}^G \hat{N}^{g}~\hat{p}_{\bullet l,ac}^g, \label{aac:est}\\
      \hat{p}_{kl,aac} = N^{-1} \sum_{g=1}^G \hat{N}^{g}~\hat{p}_{kl,ac}^g. \nonumber
    \end{eqnarray}
It can be shown that $B(\hat{p}_{\diamond,aac}) \simeq 0$ for any $\diamond \in \{k\bullet,\bullet l,kl\}$. As for the ACC estimators, the AAC estimators can be viewed as propensity score adjusted estimators, where the response probability of a unit within an imputation class is estimated by the response rate based on available respondents within the same class. Also, as for the ACC estimators, the AAC estimators will necessarily lead to impossible values.

\section{Variance estimation under the balanced procedure} \label{sec:boot}

In this section, we turn our attention to estimating the variance of the imputed estimators under the proposed balanced imputation procedure described in Section \ref{sec:imp:proc}. It is well known that treating the imputed values as if they were observed leads to serious underestimation of the variance of imputed estimators if the proportion of missing data is appreciable and to poor confidence intervals. Several variance estimation methods accounting for nonresponse and imputation have been proposed in the literature; see Haziza (2009) for a review. In this paper, we focus on the bootstrap method, which was studied by Shao and Sitter (1996). The rationale behind their method is to select, using any complete data bootstrap method, a bootstrap sample consisting of original or rescaled imputed data and their corresponding original response statuses. The bootstrap data with a missing status are then reimputed using the same imputation method that was used in the original sample. The proposed balanced imputation procedure entails the application of the procedure within each bootstrap sample, which may be highly computer intensive. A simplified bootstrap method can be used by noting that the imputation variance is virtually eliminated under the proposed balanced imputation procedure. It consists of reimputing the deterministic version of the balanced procedure within each bootstrap sample, which is equivalent to re-calculating $\tilde{p}_{\diamond,I} \equiv E_{I} \left(\hat{p}_{\diamond,I}\right)$ within each bootstrap sample, $\diamond \in \{k\bullet,\bullet l,kl\}$. After some relatively straightforward algebra, we obtain
    \begin{eqnarray}
      \tilde {p}_{k \bullet,I} & \simeq & {N}^{-1} \sum_{g=1}^G \left[\hat{N}_{r\bullet}^{g} \hat{p}_{k\bullet,ac}^g
                                                                         +\hat{N}_{mr}^{g} \hat{p}_{k\bullet,mr}^g
                                                                         +\hat{N}_{mm}^{g} \hat{p}_{k\bullet,cc}^g\right], \nonumber \\
      \tilde {p}_{\bullet l,I} & \simeq & {N}^{-1} \sum_{g=1}^G \left[\hat{N}_{\bullet r}^{g} \hat{p}_{\bullet l,ac}^g
                                                                         +\hat{N}_{rm}^{g} \hat{p}_{\bullet l,rm}^g
                                                                         +\hat{N}_{mm}^{g} \hat{p}_{\bullet l,cc}^g\right], \label{p:tilde}\\
      \tilde {p}_{kl,I} & \simeq & {N}^{-1} \sum_{g=1}^G \left[(\hat{N}_{rr}^{g}+\hat{N}_{mm}^{g}) \hat{p}_{kl,cc}^g
                                                                  +\hat{N}_{mr}^{g} \hat{p}_{kl,mr}^g +\hat{N}_{rm}^{g} \hat{p}_{kl,rm}^g \right], \nonumber
    \end{eqnarray}
where $\hat{p}_{k\bullet,ac}^g$ and $\hat{p}_{\bullet l,ac}^g$ are given in (\ref{ac:est}), $\hat{p}_{k\bullet,cc}^g$, $\hat{p}_{\bullet l,cc}^g$ and $\hat{p}_{kl,cc}^g$ are given in (\ref{cc:est}) and
    \begin{eqnarray*}
       \hat{p}_{k\bullet,mr}^g & = & \frac{\sum_{i \in s_{mr}^g} w_i \sum_{l=1}^L 1(y_i=l) \hat{p}_{k|l,cc}^g}{\sum_{i \in s_{mr}^g} w_i},\\
       \hat{p}_{\bullet l,rm}^g & = & \frac{\sum_{i \in s_{rm}^g} w_i \sum_{k=1}^K 1(x_i=k) \hat{p}_{l|k,cc}^g}{\sum_{i \in s_{rm}^g} w_i},\\
       \hat{p}_{kl,mr}^g       & = & \frac{\sum_{i \in s_{mr}^g} w_i 1(y_i=l) \hat{p}_{k|l,cc}^g}{\sum_{i \in s_{mr}^g} w_i},\\
       \hat{p}_{kl,rm}^g       & = & \frac{\sum_{i \in s_{rm}^g} w_i 1(x_i=k) \hat{p}_{l|k,cc}^g}{\sum_{i \in s_{rm}^g} w_i}.
    \end{eqnarray*}

\noindent As an illustration, we use the bootstrap weight method of Rao, Wu and Yue (1992) in the special case of simple random sampling without replacement. The extension to stratified simple random sampling without replacement is straightforward. The bootstrap weight procedure proceeds as follows:
    \begin{itemize}
      \item[(1)] Let $n'$ be the bootstrap sample size, which may be different from $n$.
      \item[(2)] Draw a simple random sample \textit{with} replacement $s^*$ of size $n'$ from $s$. Let $m_i^*$ be the number of times unit $i$ is selected in $s^*$. We have $n'=\sum_{i \in s} m_i^*$. For unit $i \in s$, define the bootstrap weight as
            \begin{eqnarray*}
              w_i^* = w_i \left\{1+\sqrt{C}\left(\frac{n m_i^*}{n'}-1\right)\right\} & \textrm{ with } & C=\frac{n'\left(1-\frac{n}{N}\right)}{n-1}.
            \end{eqnarray*}
      Compute $\tilde {p}_{\diamond,I}^*$ from (\ref{p:tilde}) by replacing $w_i$ with $w_i^*$.
      \item[(3)] Repeat Step 2 a large number of times, $C$, to get $\tilde {p}_{\diamond,I}^{*(1)},\ldots,\tilde {p}_{\diamond,I}^{*(C)}$.
      \item[(4)] Estimate $V_{p}\left(\tilde {p}_{\diamond,I}|\mathbf{r} \right)$ by
        \begin{eqnarray}\label{bootvar}
           \hat{V}_{1C} = \frac{1}{C-1} \sum_{c=1}^C \left(\tilde {p}_{\diamond,I}^{*(c)}- \frac{1}{C} \sum_{d=1}^C \tilde {p}_{\diamond,I}^{*(d)} \right)^2.
        \end{eqnarray}
    \end{itemize}
The reader is referred to Chauvet~(2007,2015) for a review of bootstrap methods in survey sampling, and to Antal and Till\'{e} (2011) and Beaumont and Patak (2012) for bootstrap weight methods in the context of unequal probability sampling designs. If the sampling fraction $n/N$ is negligible, the bootstrap variance estimators (\ref{bootvar}) are consistent for the true variance; see Haziza (2009) and Mashreghi et al. (2014) for a discussion on the consistency of the method of Shao and Sitter (1996). Variance estimation for non-negligible sampling fractions in the context of bivariate parameters requires further investigations.

\section{Simulation study} \label{sec:sim}

We conducted two simulation studies to test the performance of the point and variance estimation procedures described in Sections \ref{Sec:setup}-\ref{sec:boot}. In the first study, we compared the performance of several point estimation procedures in terms of relative bias and relative efficiency. In the second, we tested the performance of the bootstrap variance estimator described in Section \ref{sec:boot}.

\subsection{Performance of the point estimators} \label{ssec1:sim}

We generated a finite population of size $N=20,000$ consisting of two binary variables $x$ and $y$ so that $k \in \{0,1\}$ and $l \in \{0,1\}.$ The population consisted of five classes, each of size $4,000$. We were interested in estimating the marginal first moments $p_{1\bullet}$ and $p_{\bullet 1}$, the joint proportion $p_{11}$ as well as the population odd-ratio
    \begin{equation}\label{OR}
    \rm{OR} =\frac{p_{11}~p_{00}}{p_{10}~p_{01}}.
    \end{equation}

\noindent From the population, we selected $B=10,000$ samples of size $n=2,000$ according to simple random sampling without replacement. In each selected sample, non-response to $x$ and $y$ was generated according to a non-response mechanism described in Table \ref{tab:rb}, along with the population characteristics. The characteristics of the population were chosen so as to obtain a positive association between $\phi_{rr}^g$ and $p_{1 \bullet}^g$, between $\phi_{rr}^g$ and $p_{\bullet 1}^g$, and between $\phi_{rr}^g$ and $p_{11}^g$. The CC estimators are therefore expected to be positively biased; see equations (\ref{bias:cc:est}). Also, the characteristics of the population were chosen so as to obtain a positive association between $\phi_{r\bullet}^g$ and $p_{1 \bullet}^g$, and between $\phi_{\bullet r}^g$ and $p_{\bullet 1}^g$. The AC estimators are therefore expected to be positively biased; see equations (\ref{bias:ac:est}). \\

\begin{table}[htb!]
    \begin{center}
    \caption{Characteristics of the population and mechanism used to generate nonresponse \label{tab:rb}}
    \begin{tabular}{|cc|cccc|cccc|}
    \hline
    Class & & $p_{1\bullet}$ & $p_{\bullet 1}$ & $p_{11}$ & $OR$ & $\phi_{rr}$ & $\phi_{rm}$ & $\phi_{mr}$ & $\phi_{mm}$ \\
     \hline
     1  & & 0.50 & 0.50 & 0.20 & 0.44  & 0.10 & 0.20 & 0.20 & 0.50 \\
     2  & & 0.55 & 0.55 & 0.30 & 0.96  & 0.20 & 0.20 & 0.20 & 0.40 \\
     3  & & 0.60 & 0.60 & 0.40 & 2.00  & 0.30 & 0.25 & 0.25 & 0.20 \\
     4  & & 0.65 & 0.65 & 0.50 & 4.44  & 0.40 & 0.20 & 0.20 & 0.20 \\
     5  & & 0.70 & 0.70 & 0.60 & 12.00 & 0.50 & 0.20 & 0.20 & 0.10 \\ \hline
    \end{tabular}
    \end{center}
    \end{table}

\noindent In each sample, we computed seven estimators for each of the parameters of interest $p_{k\bullet},$ $ p_{\bullet l},$  $p_{11}$ and OR: (i) the CC estimators given in equations (\ref{cc:est}); (ii) the ACC estimators given in equations (\ref{acc:est}); (iii) the AC estimators given in equations (\ref{ac:est}); (iv) the AAC estimators given in equations (\ref{aac:est}); (v) the imputed estimators given by (\ref{est:imp}) based on the random hot-deck imputation (RHDI) procedure described in Section \ref{Sec:setup}; (vi) the imputed estimators given by (\ref{est:imp}) based on the joint random hot-deck imputation (JHDI) procedure described in Section \ref{sec:imp:proc}; (vii) the imputed estimators given by (\ref{est:imp}) based on the balanced random hot-deck imputation (BHDI) procedure described in Section \ref{sec:imp:proc}. In each case, an estimator $\widehat{\rm{OR}}_I$ of the OR was obtained by replacing each unknown parameter in (\ref{OR}) by its corresponding imputed estimator. \\

\noindent As a measure of bias of a point estimator $\hat{\theta}$ of a parameter $\theta$, we used the Monte Carlo Percent Relative Bias ($RB$) given by
    \begin{equation} \label{rb}
      RB(\hat{\theta})=\frac{E_{MC}(\hat{\theta})-\theta}{\theta} \times 100,
    \end{equation}
where $E_{MC}(\hat{\theta})=B^{-1}\sum_{b=1}^B \hat{\theta}^{(b)}$ and $\hat{\theta}^{(b)}$ denotes the estimator $\hat{\theta}$ in the $b$-th sample, $b=1,\ldots,10~000$. When the true value of the parameter $\theta$ is close to zero, the relative bias may not be an appropriate measure. This is not problematic in our simulation set-up as the values of $p_{1\bullet}$, $p_{\bullet 1}$, $p_{11}$ and $OR$ were bounded away from 0 (see Table \ref{tab:rb}). As a measure of Relative Efficiency (RE), we used
    \begin{equation} \label{re}
      RE=\frac{MSE_{MC}(\hat{\theta}^{(AAC)})}{MSE_{MC}(\hat{\theta}^{(\cdot)})} \times 100,
    \end{equation}
where $MSE_{MC}(\hat{\theta})$ is the Monte Carlo mean square error of $\hat \theta$ and $\hat{\theta}^{AAC}$  denote the adjusted available-case estimator. \\

  \begin{table}[htb!]
    \begin{center}
    \caption{Monte-Carlo percent relative bias and relative efficiency (between brackets) of several estimators \label{tab:rb:re}}
    \begin{tabular}{|c c|c c c c|}
    \hline
    Estimator & & $p_{1\bullet}$ & $p_{\bullet 1}$ & $p_{11}$ & $OR$ \\
    \hline
    CC   & & 5.6 (15) &  5.5 (17) & 16.7 (10) & 71.2  (28)  \\
    ACC  & & 0.0 (46) &  0.0 (44) & 0.0 (100) & 35.6  (100) \\
    AC   & & 3.3 (41) &  3.3 (42) & 16.7 (10) & 71.2  (28)  \\
    AAC  & & 0.0 (100)&  0.0 (100)& 0.0 (100) & 35.6  (100) \\
    RHDI & & 0.0 (68) &  0.0 (68) & -3.7 (89) & -21.8 (278) \\
    JHDI & & 0.0 (60) &  0.0 (59) & 0.0 (115) & 2.5   (329) \\
    BHDI & & 0.0 (70) &  0.0 (67) & 0.0 (131) & 2.3   (377) \\
     \hline
    \end{tabular}
    \end{center}
    \end{table}

\noindent Table \ref{tab:rb:re} shows the Monte Carlo percent Relative Bias (RB) and percent Relative Efficiency (RE) of the seven estimators of $p_{1\bullet}$, $p_{\bullet 1}$, $p_{11}$ and $OR$. The CC estimators and the AC estimators showed positive bias for $p_{1\bullet}$, $p_{\bullet 1}$ and $p_{11}$, as expected. As a result, the corresponding estimators of OR were strongly biased with a value of RB equal to $71.2 \%$. The ACC estimator and the AAC estimator, which account for class information, showed virtually no bias for $p_{1\bullet}$, $p_{\bullet 1}$ and $p_{11}$, but were significantly biased for $OR$ with a value of RB equal to $35.6\%$. Turning to the imputed estimators, we note that the imputed estimators of the marginal proportions showed no bias, as expected. However, under RHDI, both the imputed estimator of $p_{11}$ and the estimator of OR were biased with values of RB equal to $-3.7\%$ and $-21.6\%,$ respectively. Also, the biases were negative clearly illustrating the problem of attenuation of relationships. On the other hand, both JHDI and BHDI led to negligible bias, showing that both procedures succeeded in preserving the relationship between variables.\\

\noindent We now turn to the relative efficiency. We first consider the marginal first moments. We note that the CC and ACC estimators were inefficient, which can be explained by the fact that they tend to discard a lot of information. The imputed estimators under both RHDI and JHDI were less efficient than the corresponding AAC estimator with values of RE ranging from $59\%$ to $68\%$. The efficiency loss arises from the random selection of donors in the random hot-deck imputation procedures. The imputed estimators under BHDI were more efficient than the corresponding estimators obtained under RHDI and JHDI, illustrating the reduction of the imputation variance. In regards to the joint proportion $p_{11}$, the imputed estimator under RHDI was less efficient than the AAC estimators, while the imputed estimators under both JHDI and BHDI were more efficient. The imputed estimator of $OR$ under all three imputation methods was considerably more efficient than the AAC estimator.

\subsection{Performance of the variance estimators}

We conducted a second simulation study on the same population in order to assess the performance of the bootstrap procedure described in Section \ref{sec:boot}. We were interested in estimating the variance of the marginal first moments $p_{1\bullet}$ and $p_{\bullet 1}$, the joint proportion $p_{11}$ as well as the population odd-ratio $\rm{OR}$. \\

\noindent From each population, we selected $B=10,000$ samples of size $n=1,000$ according to simple random sampling without replacement. In each selected sample, non-response to $x$ and $y$ was generated according to the non-response mechanism described in Table \ref{tab:rb}. We were interested in estimating the variance of the imputed estimators of $p_{1\bullet}$, $p_{\bullet 1}$, $p_{11}$ and $\rm{OR}$ under the proposed balanced random imputation. In each sample (containing respondents and nonrespondents), we selected $B=2,000$ bootstrap samples according to the bootstrap weight procedure of Section \ref{sec:boot}. To measure the bias of the Bootstrap variance estimator, we used the Monte Carlo percent relative bias given by (\ref{rb}). The true variance was replaced by a Monte Carlo approximation, obtained through an independent run of $50,000$ simulations. Also, we computed confidence intervals by means of the percentile method. For example, in the case of $\widehat{\rm{OR}}_{I}$, we computed the $B$ bootstrap versions of the odd-ratio, $\tilde{\rm{OR}}_{I}^{*(b)} ,b=1,\ldots,B$. An $(1-2\alpha)$ confidence interval is then given by $\left[\tilde{\rm{OR}}_{I}^{*(L)},\tilde{\rm{OR}}_{I}^{*(U)} \right]$ with $L=\alpha~B$ and $U=(1-\alpha)~B$. Error rates of the confidence intervals (with nominal error rates of $2.5 \%\ $ and $5 \%\ $ in each tail) were compared. \\

\noindent Table \ref{rBias} shows the Monte Carlo percent relative bias (RB) of the Bootstrap variance estimator and the error rates. The Bootstrap variance estimator performed well for $\hat{p}_{1\bullet,I}$, $\hat{p}_{\bullet 1,I}$ and $\hat{p}_{11,I}$, with an absolute relative bias less than $5 \%$. The Bootstrap variance estimator was positively biased for $\widehat{\rm{OR}}_{I}$. The error rates were close to the nominal rates in all the cases.

    \begin{table}[htb!]
    \caption{Monte Carlo percent RB (in \%) and error rates of the Bootstrap variance estimator} \label{rBias}
    \begin{center}
    \begin{tabular}{|c|c|ccc|ccc|}
    \hline
    & RB & \multicolumn{3}{c|}{Coverage rate 2.5 \%\ } & \multicolumn{3}{c|}{Coverage rate 5 \%\ } \\ \cline{3-5} \cline{6-8}
    & & L & U & L+U & L & U & L+U \\ \hline \hline
    $\hat{p}_{1 \bullet,I}$ & -3.9 & 2.9 & 3.4 & 6.3 & 5.2 & 5.7 & 10.9 \\
    $\hat{p}_{\bullet 1,I}$ & -5.0 & 3.4 & 3.9 & 7.3 & 5.9 & 6.4 & 12.3 \\
    $\hat{p}_{11,I}$        & -3.9 & 2.5 & 3.4 & 5.9 & 5.6 & 6.1 & 11.7 \\
    $\rm{OR}_{I}$           & 16.2 & 3.2 & 3.3 & 6.5 & 5.2 & 5.8 & 11.0 \\
    \hline
    \end{tabular}
    \end{center}
    \end{table}


\section{Concluding remarks}

In this paper, we considered the problem of preserving the relationship between categorical variables when imputation was used to compensate for the missing values. We proposed a simple joint imputation procedure that succeeds in preserving the relationship between two categorical variables, unlike random hot-deck imputation. We also proposed a fully efficient version of the proposed joint imputation procedure. Simulation results showed the good performance of both methods in terms of bias. Also, the balanced random hot-deck imputation procedure was found to be significantly more efficient than the joint random hot-deck imputation procedure.

\section*{References}

\noindent Andridge, R.R. and Little, R.J.A (2010). A review of hot deck imputation for survey non-response. \emph{International Statistical Review}, 78, 40--64. \\

\noindent Antal, E., and Till\'e, Y. (2011). A direct bootstrap method for complex sampling designs from a finite population. \textit{Journal of the American Statistical Association}, 106, 534--543. \\

\noindent Beaumont, J.-F., and Patak, Z. (2012). On the generalized bootstrap for sample surveys with special attention to poisson sampling. \textit{International Statistical Review} 80, 127--148. \\

\noindent Chauvet, G. (2007). M\'ethodes de Bootstrap en population finie. \textit{Ph.D. dissertation}, University of Rennes 2. \\

\noindent Chauvet, G. (2015). Coupling methods for multistage sampling. \textit{Annals of Statistics}, to appear. \\

\noindent Chauvet, G., Deville, J.C., and Haziza, D. (2011). On Balanced Random Imputation in Surveys. \emph{Biometrika}, 98, 459--471. \\

\noindent Chauvet, G., and Haziza, D. (2012). Fully efficient estimation of coefficients of correlation in the presence of imputed data. \emph{The Canadian Journal of Statistics}, 40, 124--149. \\

\noindent Deville, J.C. (1999). Variance estimation for complex statistics and estimators: linearization and residual techniques. \emph{Survey Methodology},  25, 193--203. \\

\noindent Deville, J.C. and Till\'e, Y. (2004). Efficient balanced sampling: the cube method. \emph{Biometrika},  91, 893--912. \\

\noindent Fay,~R.~E. (1991). A Design-Based Perspective on Missing Data Variance. \emph{Proceedings of the 1991 Annual Research Conference, US Bureau of the Census},  429--440. \\

\noindent Hajek, J. (1971) Comment on An essay on the logical foundations of survey sampling by Basu, D. in Foundations of Statistical Inference (Godambe, V.P. and Sprott, D.A. eds.), p. 236. Holt, Rinehart and Winston. \\

\noindent Haziza, D. and Beaumont, J-F. (2007). On the construction of imputation classes in surveys. \emph{International Statistical Review}, 75,  25--43. \\

\noindent Haziza,~D. (2009). Imputation and inference in the presence of missing data. \emph{Handbook of Statistics, Volume 29, Sample Surveys: Theory Methods and Inference}, Editors: C.R. Rao and D. Pfeffermann, 215--246. \\

\noindent Horvitz, D.G., and Thompson, D.J. (1952). A generalization of sampling without replacement from a finite universe. \textit{Journal of the American Statistical Association}, 47, 663--685.  \\

\noindent Kim, J.K. and Fuller, W.A. (2004). Fractional hot-deck imputation. \emph{Biometrika}, 91, 559--578.  \\

\noindent Mashreghi, Z., L\'{e}ger, C. and Haziza, D. (2014). Bootstrap Methods for Imputed Data from Regression, Ratio and Hot Deck Imputation. \emph{The Canadian Journal of Statistics}, 42, 162--167. \\

\noindent Rubin, D.B. (1976). Inference and Missing Data. \emph{Biometrika}, 63, 581--590. \\

\noindent Shao, J. and Wang, H. (2002). Sample correlation coefficients based on survey data under regression imputation. \emph {Journal of the American Statistical Association},  97, 544--552. \\

\noindent Shao,~J. and Steel,~P. (1999). Variance estimation for survey data with composite imputation and nonnegligible sampling fractions. \emph{Journal of the American Statistical Association}, 94, 254--265. \\

\noindent Shao,~J. and Tu,~D. (1995). \emph{The jackknife and the bootstrap}. Springer. \\

\noindent Skinner,~C.~J. and Rao,~J.~N.~K. (2002). Jackknife variance for multivariate statistics under hot deck imputation from common donors. \emph{Journal of Statistical Planning and Inference}, 102, 149--167.

\newpage
\appendix

\section{Proof of equation (\ref{BqI:randHD})} \label{sec:proof:BqI:randHD}

From the definition of $\hat{p}_{k\bullet,I}$, we have
    \begin{eqnarray*}
      E_I \left(\hat{p}_{k\bullet,I}\right) & = & N^{-1} \sum_{g=1}^G \sum_{i \in s_{r\bullet}^g} w_i 1(x_i=k) \\
                                            & + & N^{-1} \sum_{g=1}^G \sum_{i \in s_{mr}^g} w_i \hat{p}_{k \bullet,ac}^g+N^{-1} \sum_{g=1}^G \sum_{i \in s_{mm}^g} w_i \sum_{l=1}^L \hat{p}_{kl,cc}^g .
    \end{eqnarray*}
Since $E_q(\hat{p}_{k \bullet,ac}^g) \simeq \hat{p}_{k \bullet}^g$ and $E_q(\sum_{l=1}^L \hat{p}_{kl,cc}^g) \simeq \hat{p}_{k \bullet}^g$, we obtain
    \begin{eqnarray*}
      E_{qI} \left(\hat{p}_{k\bullet,I}\right) & \simeq & N^{-1} \sum_{g=1}^G \sum_{i \in s^g} w_i (\phi_{rr}^g+\phi_{rm}^g) 1(x_i=k) \\
                                                 & + & N^{-1} \sum_{g=1}^G \hat{p}_{k \bullet}^g \sum_{i \in s^g} w_i \phi_{mr}^g +N^{-1} \sum_{g=1}^G \hat{p}_{k \bullet}^g \sum_{i \in s^g} w_i \phi_{mm}^g \\
                                                 & = & N^{-1} \sum_{g=1}^G (\phi_{rr}^g+\phi_{rm}^g) \sum_{i \in s^g} w_i 1(x_i=k)
                                                 + N^{-1} \sum_{g=1}^G (\phi_{mr}^g+\phi_{mm}^g) \sum_{i \in s^g} w_i 1(x_i=k) \\
                                                 & = & N^{-1} \sum_{i \in s} w_i 1(x_i=k),
    \end{eqnarray*}
so that $B_{qI} \left(\hat{p}_{k\bullet,I}\right) \simeq 0$. The proof for $\hat{p}_{\bullet l,I}$ is similar. We now turn to $\hat{p}_{kl,I}$. From definition, we have
    \begin{eqnarray*}
      E_I(\hat{p}_{kl,I}) & = & N^{-1} \sum_{g=1}^G \sum_{i \in s_{rr}^g} w_i 1(x_i=k) 1(y_i=l) + N^{-1} \sum_{g=1}^G \sum_{i \in s_{rm}^g} w_i 1(x_i=k) \hat{p}_{\bullet l,ac}^g \\
                              & + & N^{-1} \sum_{g=1}^G \sum_{i \in s_{mr}^g} w_i \hat{p}_{k \bullet,ac}^g 1(y_i=l) + N^{-1} \sum_{g=1}^G \sum_{i \in s_{mm}^g} w_i \hat{p}_{kl,cc}^g \\
                              & = & N^{-1} \sum_{g=1}^G \sum_{i \in s_{rr}^g} w_i 1(x_i=k) 1(y_i=l) + N^{-1} \sum_{g=1}^G \hat{p}_{\bullet l,ac}^g \sum_{i \in s_{rm}^g} w_i 1(x_i=k) \\
                              & + & N^{-1} \sum_{g=1}^G \hat{p}_{k \bullet,ac}^g \sum_{i \in s_{mr}^g} w_i 1(y_i=l) + N^{-1} \sum_{g=1}^G \hat{p}_{kl,cc}^g \sum_{i \in s_{mm}^g} w_i .
    \end{eqnarray*}
Since $E_q(\hat{p}_{\bullet l,ac}^g) \simeq \hat{p}_{\bullet l}^g$, $E_q(\hat{p}_{k \bullet,ac}^g) \simeq \hat{p}_{k \bullet}^g$ and $E_q(\hat{p}_{kl,cc}^g) \simeq \hat{p}_{kl}^g$, we obtain
    \begin{eqnarray*}
      E_{qI}(\hat{p}_{kl,I}) & \simeq & N^{-1} \sum_{g=1}^G \phi_{rr}^g \sum_{i \in s^g} w_i 1(x_i=k) 1(y_i=l) + N^{-1} \sum_{g=1}^G \hat{p}_{\bullet l}^g \times \phi_{rm}^g \sum_{i \in s^g} w_i 1(x_i=k) \\
                               & + & N^{-1} \sum_{g=1}^G \hat{p}_{k \bullet}^g \times \phi_{mr}^g \sum_{i \in s^g} w_i 1(y_i=l) + N^{-1} \sum_{g=1}^G \hat{p}_{kl}^g \times \phi_{mm}^g \sum_{i \in s^g} w_i \\
                               & = & N^{-1} \sum_{g=1}^G (\phi_{rr}^g+\phi_{mm}^g) \sum_{i \in s^g} w_i 1(x_i=k) 1(y_i=l) \\
                               & + & N^{-1} \sum_{g=1}^G (\hat{N}^g)^{-1} (\phi_{rm}^g+\phi_{mr}^g) \left\{\sum_{i \in s^g} w_i 1(x_i=k)\right\} \left\{\sum_{j \in s^g} w_j 1(y_j=l)\right\},
    \end{eqnarray*}
This leads to
    \begin{eqnarray*}
      E_{qI}(\hat{p}_{kl,I}-\hat{p}_{kl}) & = & N^{-1} \sum_{g=1}^G (\hat{N}^g)^{-1} (\phi_{rm}^g+\phi_{mr}^g) \left\{\sum_{i \in s^g} w_i 1(x_i=k)\right\} \left\{\sum_{j \in s^g} w_j 1(y_j=l)\right\} \\
                                            & - & N^{-1} \sum_{g=1}^G (\phi_{rm}^g+\phi_{mr}^g) \sum_{i \in s^g} w_i 1(x_i=k) 1(y_i=l) \\
                                            & = & - N^{-1} \sum_{g=1}^G (\phi_{rm}^g+\phi_{mr}^g) \sum_{i \in s^g} w_i \{1(x_i=k)-\hat{p}_{k\bullet}^g\} \{1(y_i=l)-\hat{p}_{\bullet l}^g\},
    \end{eqnarray*}
and
    \begin{eqnarray*}
      E_{pqI}(\hat{p}_{kl,I}-\hat{p}_{kl}) & \simeq & - N^{-1} \sum_{g=1}^G (\phi_{rm}^g+\phi_{mr}^g) \sum_{i \in U^g} \{1(x_i=k)-p_{k\bullet}^g\} \{1(y_i=l)-p_{\bullet l}^g\},
    \end{eqnarray*}
which leads to (\ref{BqI:randHD}).

\section{Non-response bias for the imputed estimators under the proposed procedures} \label{appen:B}

We first consider $\hat{p}_{k\bullet,I}$. From definition, we have
    \begin{eqnarray}
      E_I \left(\hat{p}_{k\bullet,I}\right) & = & N^{-1} \sum_{g=1}^G \sum_{i \in s_{r\bullet}^g} w_i 1(x_i=k) \label{EI:pkI:jhdi}\\
                                            & + & N^{-1} \sum_{g=1}^G \sum_{i \in s_{mr}^g} w_i \sum_{l=1}^L 1(y_i=l) \hat{p}_{k|l,cc}^g+N^{-1} \sum_{g=1}^G \sum_{i \in s_{mm}^g} w_i \sum_{l=1}^L \hat{p}_{kl,cc}^g. \nonumber
    \end{eqnarray}
Since $E_q(\hat{p}_{k|l,cc}^g) \simeq \frac{\hat{p}_{kl}^g}{\hat{p}_{\bullet l}^g}$ and $E_q(\sum_{l=1}^L \hat{p}_{kl,cc}^g) \simeq \hat{p}_{k \bullet}^g$, we obtain
    \begin{eqnarray*}
      E_{qI} \left(\hat{p}_{k\bullet,I}\right) & \simeq & N^{-1} \sum_{g=1}^G \sum_{i \in s^g} w_i (\phi_{rr}^g+\phi_{rm}^g) 1(x_i=k) \\
                                                 & + & N^{-1} \sum_{g=1}^G \sum_{i \in s^g} w_i \phi_{mr}^g \sum_{l=1}^L 1(y_i=l) \frac{\hat{p}_{kl}^g}{\hat{p}_{\bullet l}^g} + N^{-1} \sum_{g=1}^G \hat{p}_{k \bullet}^g \sum_{i \in s^g} w_i \phi_{mm}^g \\
                                                 & = & N^{-1} \sum_{i \in s} w_i 1(x_i=k)=\hat{p}_{k\bullet},
    \end{eqnarray*}
so that $B_{pqI} \left(\hat{p}_{k\bullet,I}\right) \simeq 0$. The proof for $\hat{p}_{\bullet l,I}$ is similar. We now turn to $\hat{p}_{kl,I}$. Using similar arguments, we obtain
    \begin{eqnarray}
      E_I(\hat{p}_{kl,I}) & = & N^{-1} \sum_{g=1}^G \sum_{i \in s_{rr}^g} w_i 1(x_i=k) 1(y_i=l) + N^{-1} \sum_{g=1}^G \sum_{i \in s_{rm}^g} w_i 1(x_i=k) \hat{p}_{l|k,cc}^g \nonumber \\
                          & + & N^{-1} \sum_{g=1}^G \sum_{i \in s_{mr}^g} w_i 1(y_i=l) \hat{p}_{k|l,cc}^g + N^{-1} \sum_{g=1}^G \sum_{i \in s_{mm}^g} w_i \hat{p}_{kl,cc}^g \label{EI:pklI:jhdi}
    \end{eqnarray}
and
    \begin{eqnarray*}
      E_{qI}(\hat{p}_{kl,I}) & \simeq & N^{-1} \sum_{g=1}^G \phi_{rr}^g \sum_{i \in s^g} w_i 1(x_i=k) 1(y_i=l)) + N^{-1} \sum_{g=1}^G \phi_{rm}^g \sum_{i \in s^g} w_i 1(x_i=k) \frac{\hat{p}_{kl}^g}{\hat{p}_{k\bullet}^g} \\
                             & + & N^{-1} \sum_{g=1}^G \phi_{mr}^g \sum_{i \in s^g} w_i 1(y_i=l) \frac{\hat{p}_{kl}^g}{\hat{p}_{\bullet l}^g} + N^{-1} \sum_{g=1}^G \phi_{mm}^g \sum_{i \in s^g} w_i \hat{p}_{kl}^g \\
                             & = & N^{-1} \sum_{i \in s} w_i 1(x_i=k) 1(y_i=l) = \hat{p}_{kl},
    \end{eqnarray*}
so that $B_{pqI} \left(\hat{p}_{kl,I}\right) \simeq 0$.

\section{Extension of the proposed imputation procedures} \label{sec:extension}

In this section, we briefly describe the set-up and extension of the imputation procedures to the case of more than two missing items. To avoid intricate notations, we focus on the case of $3$ missing items and describe the extension of the joint random hot-deck imputation only. In addition to $x$ and $y,$ let $z$ denote a study variable with $Q$ possible characteristics $z_i=0,\ldots,Q-1$ for unit $i$. We want to impute jointly the three variables $x$, $y$ and $z$. We assume that the population $U$ is partitioned into $G$ imputation classes $U_1,\ldots,U_G$ and note $s_{\circ}^g$ the subset of units in $s^g=S \cap U^g$ with pattern $\circ \in \{rrr,mrr,rmr,rrm,mmr,mrm,rmm,mmm\}$, where the first letter in $\circ$ refers to the status of  $x$ (respondent or missing), the second to the status of $y$ and the third to the status of $z$. We assume that the data are MCAR within imputation classes, and we note $P(i \in s_{\diamond}^g | i \in s)=\phi_{\diamond}^g$.

The joint random imputation procedure described in Section \ref{sec:imp:proc} can be extended by modeling the distribution of each variable conditionally on the non-missing items known for this variable. For any unit $i \in U^g$; we note
    \begin{eqnarray*}
      \hat{p}_{k|lq,cc}^g & = & \frac{\sum_{i \in s_{rrr}^g} w_i 1(x_i=k) 1(y_i=l) 1(z_i=q)}{\sum_{i \in s_{rrr}^g} w_i 1(y_i=l) 1(z_i=q)}, \\
      \hat{p}_{l|kq,cc}^g & = & \frac{\sum_{i \in s_{rrr}^g} w_i 1(x_i=k) 1(y_i=l) 1(z_i=q)}{\sum_{i \in s_{rrr}^g} w_i 1(x_i=k) 1(z_i=q)}, \\
      \hat{p}_{q|kl,cc}^g & = & \frac{\sum_{i \in s_{rrr}^g} w_i 1(x_i=k) 1(y_i=l) 1(z_i=q)}{\sum_{i \in s_{rrr}^g} w_i 1(x_i=k) 1(y_i=l)},
    \end{eqnarray*}
for the estimated conditional probabilities when two items are available; we note
    \begin{eqnarray*}
      \hat{p}_{kl|q,cc}^g & = & \frac{\sum_{i \in s_{rrr}^g} w_i 1(x_i=k) 1(y_i=l) 1(z_i=q)}{\sum_{i \in s_{rrr}^g} w_i 1(z_i=q)}, \\
      \hat{p}_{kq|l,cc}^g & = & \frac{\sum_{i \in s_{rrr}^g} w_i 1(x_i=k) 1(y_i=l) 1(z_i=q)}{\sum_{i \in s_{rrr}^g} w_i 1(y_i=l)}, \\
      \hat{p}_{lq|k,cc}^g & = & \frac{\sum_{i \in s_{rrr}^g} w_i 1(x_i=k) 1(y_i=l) 1(z_i=q)}{\sum_{i \in s_{rrr}^g} w_i 1(x_i=k)},
    \end{eqnarray*}
for the estimated conditional probabilities when one item is available; finally, we note
    \begin{eqnarray*}
      \hat{p}_{klq,cc}^g & = & \frac{\sum_{i \in s_{rrr}^g} w_i 1(x_i=k) 1(y_i=l) 1(z_i=q)}{\sum_{i \in s_{rrr}^g} w_i}.
    \end{eqnarray*}

The joint random imputation procedure is as follows:
\begin{itemize}
\item [(i)] for $i \in s_{mrr}^g$, missing $x_{i}$ is imputed by $x_{i}^{*}=k$ with probability $\hat{p}_{k|y_i z_i,cc}^g$;
\item [(ii)] for $i \in s_{rmr}^g$, missing $y_{i}$ is imputed by $y_{i}^{*}=l$ with probability $\hat{p}_{l|x_i z_i,cc}^g$;
\item [(iii)] for $i \in s_{rrm}^g$, missing $z_{i}$ is imputed by $z_{i}^{*}=q$ with probability $\hat{p}_{q|x_i y_i,cc}^g$;
\item [(iv)] for $i \in s_{mmr}^g$, missing $(x_i,y_{i})$ is imputed by $(x_i^*,y_{i}^{*})=(k,l)$ with probability $\hat{p}_{kl|z_i,cc}^g$;
\item [(v)] for $i \in s_{mrm}^g$, missing $(x_i,z_{i})$ is imputed by $(x_i^*,z_{i}^{*})=(k,q)$ with probability $\hat{p}_{kq|y_i,cc}^g$;
\item [(vi)] for $i \in s_{rmm}^g$, missing $(y_i,z_{i})$ is imputed by $(y_i^*,z_{i}^{*})=(l,q)$ with probability $\hat{p}_{lq|x_i,cc}^g$;
\item [(vii)]for $i \in s_{mmm}^g$, missing $\left(x_{i},y_{i},z_i\right)$ is imputed by $\left(x_{i}^{*},y_{i}^{*},z_i^*\right)=\left(k,l,q\right)$ with probability $\hat{p}_{klq,cc}^g$.
\end{itemize}

\section{Properties of the balanced procedure} \label{appen:C}

In this Section, we prove that $\hat{p}_{\diamond,I} = \tilde{p}_{\diamond,I}$ for $\diamond \in \{k'\bullet,\bullet l',k'l'\}$ and any characteristics $k'$ and $l'$. We first consider $\hat{p}_{k' \bullet,I}$ for $k'=1,\ldots,K$. The case of $\hat{p}_{\bullet l',I}$ for $l'=1,\ldots,L$ may be proved similarly. Using (\ref{EI:pkI:jhdi}), we obtain after some algebra that a sufficient condition for $\hat{p}_{k' \bullet,I}=\tilde{p}_{k' \bullet,I}$ is that for any $g=1,\ldots,G$:
    \begin{eqnarray}
      \sum_{i \in s_{mr}^g} w_i 1(x_i^*=k') & = & \sum_{i \in s_{mr}^g} w_i \hat{p}_{k'|y_i,cc}^g, \label{app3:eq1} \\
      \sum_{i \in s_{mm}^g} w_i 1(x_i^*=k') & = & \sum_{i \in s_{mm}^g} w_i \left(\sum_{l'=1}^L \hat{p}_{k'l',cc}^g \right). \label{app3:eq2}
    \end{eqnarray}
In equation (\ref{app3:eq1}), the first term may be rewritten as
    \begin{eqnarray*}
      \sum_{i \in s_{mr}^g} w_i 1(x_i^*=k') & = & \sum_{(i,k) \in s_{mr}^{g*}} w_i 1(k=k') \sum_{l'=1}^L 1(y_i=l') \\
                                            & = & \sum_{(i,k) \in s_{mr}^{g*}} \left(\hat{p}_{k|y_i,cc}^g\right)^{-1} \left\{\sum_{l'=1}^L (\mathbf{t}_{ik})_{(k'-1)L+l'} \right\},
    \end{eqnarray*}
and the second term may be rewritten as
    \begin{eqnarray*}
      \sum_{i \in s_{mr}^g} w_i \hat{p}_{k'|y_i,cc}^g & = & \sum_{i \in s_{mr}^g} \sum_{k=1}^K w_i \hat{p}_{k|y_i,cc}^g 1(k=k') \sum_{l'=1}^L 1(y_i=l') \\
                                                      & = & \sum_{(i,k) \in U_{mr}^{g*}} \left\{\sum_{l'=1}^L (\mathbf{t}_{ik})_{(k'-1)L+l'} \right\}
    \end{eqnarray*}
so that (\ref{app3:eq1}) follows from (\ref{bal:eq:1}). Similarly, equation (\ref{app3:eq2}) follows from (\ref{bal:eq:3}). We now consider $\hat{p}_{k'l',I}$ for $k'=1,\ldots,K$ and $l'=1,\ldots,L$. Using (\ref{EI:pklI:jhdi}), we obtain after some algebra that a sufficient condition for $\hat{p}_{k'l',I}=\tilde{p}_{k'l',I}$ is that for any $g=1,\ldots,G$:
    \begin{eqnarray}
      \sum_{i \in s_{mr}^g} w_i 1(x_i^*=k')1(y_i=l') & = & \sum_{i \in s_{mr}^g} w_i \hat{p}_{k'|l',cc}^g 1(y_i=l'), \label{app3:eq3} \\
      \sum_{i \in s_{rm}^g} w_i 1(x_i=k')1(y_i^*=l') & = & \sum_{i \in s_{rm}^g} w_i \hat{p}_{l'|k',cc}^g 1(x_i=k'), \label{app3:eq4} \\
      \sum_{i \in s_{mm}^g} w_i 1(x_i^*=k')1(y_i^*=l') & = & \sum_{i \in s_{mm}^g} w_i \hat{p}_{k'l',cc}^g. \label{app3:eq5}
    \end{eqnarray}
It is easily proved that equations (\ref{bal:eq:1}), (\ref{bal:eq:2}) and (\ref{bal:eq:3}) imply equations (\ref{app3:eq3}), (\ref{app3:eq4}) and (\ref{app3:eq5}), respectively.

\end{document}